# Conception of a management tool of Technology Enhanced Learning Environments

Sérgio André Ferreira
School of Education and Psychology
Universidade Católica Portuguesa
Porto, Portugal

António Andrade
School of Economics and Management
Universidade Católica Portuguesa
Porto, Portugal

*Abstract*—This paper describes the process of the conception of a software tool of TELE management. The proposed management tool combines information from two sources: i) the automatic reports produced by the Learning Content Management System (LCMS) Blackboard and ii) the views of students and teachers on the use of the LCMS in the process of teaching and learning. The results show that the architecture of the proposed management tool has the features of a management tool, since its potential to control, to reset and to enhance the use of an LCMS in the process of teaching and learning and teacher training, is shown.

## I. INTRODUCTION

The introduction of change and educational innovation through technology in Higher Education Institutions (HEI) is a hot topic in research and in policies of international organizations and countries [1, 2]. The Learning Management Systems (LMS) and the Learning Content Management Systems (LCMS) are the most visible faces of the penetration of technology in HEI, and in many cases they are the only technological platforms to support the training activity, institutionalized and of common use by the different agents of the organization. These technological platforms are often associated with a financial investment, whose cost/ benefit ratio has to be justified.

The management of these new learning environments is critical to their success. The main factors that might endanger the success of any initiative in the introduction of technology in organizations have been identified. Several of these factors are related to aspects of management, including: introduction of technology without strategy, incipient evaluation, weak involvement of decision makers, attitudes of resistance [3].

In this study, based on the case of the Universidade Católica Portuguesa - Porto Regional Center (Católica - Porto), the process of the conception of a software tool of TELE management is described. During the school year 2003/2004, this University introduced the Blackboard LMS to support classroom teaching and to offer distance learning courses. In 2011/2012 a new investment was made for the provision of an LCMS, also Blackboard. These investments have a significant financial impact and are expected to provide a return in the educational field.

In a previous study, it was concluded that the current statistical reports produced by Blackboard neither provide critical information, nor provide a degree of disaggregation that allows the positioning of each CU, department and school/ university on the levels of integration of the LMS in the learning process [4]. The limitations of the reports affect their role as management tools, as they do not favor the dissemination of good practices or the removal of barriers, hence resulting in a slower penetration of the new culture [5, 6]. The goal of this article is to devise a management tool of the TELE that allows the possibility to give an answer to these limitations.

Planning a management tool requires a clear definition of the desired future, identifying the information subsystems required for this. In other words, it must meet the information needs of the organization and users. A tool developed without proper planning will result in a high degree of dissatisfaction among its users and will fall into disuse [7].

The approach to this problem was made by adopting a methodology of action research type, in which the researchers are actively involved in the cause of the research [8].The researchers, as users of Católica's TELE - Porto, have identified gaps in the information provided by the LCMS reports and have contributed to the information needs of the organization and users. This contribution was the basis of the work for which the representative of Blackboard/ service provider would introduce the changes required in the reports.

In addition to the objective data of the automatic LCMS reports, it is essential for management to obtain information about the users' views on the various dimensions of the TELE. This justifies the development of two questionnaires (one for teachers and one for students), so that this information could be compared with the data from the LCMS reports.

Beyond this introduction, this paper is divided into four chapters: in chapter 2 the TELE of Católica - Porto is contextualized and some data that demonstrates the dynamics of the use of the institutional TELE are presented; in chapter 3 the methodological approach is clarified and the information requirements of the organization and end-users that the management tool must address are explained; in chapter 4 the description of the whole process of the conception of the tool is presented; in chapter 5 the conclusions and proposals for future work are presented.

## II. DYNAMICS OF CATÓLICA - PORTO'S TELE

Today's society, based on information and knowledge, requires a transformation with regard to educational infrastructure. Adopting an innovative approach to the education system is mainly related to the incorporation of technology in teaching practice [9]. Fig. 1 shows a possible architecture of a TELE. Currently, the learning environment is a result of the institutional vision associated with the personal vision of each student, hence a Hybrid Institutional Personal Learning Environment. The students' Personal Learning Environment (PLE) consists of the exploration of a multiplicity of skills available in the Cloud Learning Environment, which can go beyond the institutional vision. In fact, learning takes place increasingly through social media, institutional communities, exploring web tools, libraries of digital resources, repositories of Learning Objects, and other environments, tools and resources, which together result in the construction of the student's PLE student outside of the HEI.

The institutional environment will gradually incorporate this new way of learning, strongly supported by technology. Successful experiences with the integration of web 2.0 tools in contexts as unchanging and conservative in the use of technology as are lectures to dozens of students in auditoriums [10]; the ubiquitous presence of computer labs; the use of notebooks, netbooks, tablets and smartphones for teaching purposes; the availability of CU online via LCMS, are examples of reliable indicators of the construction of TELE fostered by the institution. It is, therefore, noticed a growing interest in the integration of the technological component in the educational field.

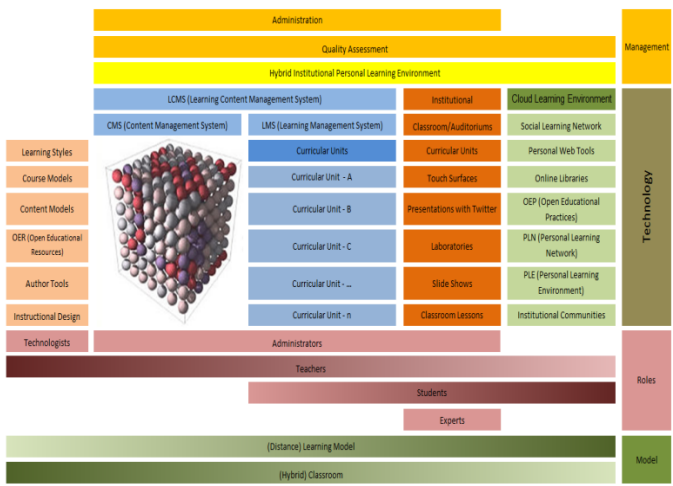

Figure 1. A possible architecture of a Hybrid Institutional Personal Learning Environment

Fig. 2 shows the LCMS as a central element of the campus of Católica - Porto, integrating the services of Learning System, Community System and Content System and maintaining communication with the administrative services – The Sophia Academic Management. The campus, supported by Windows Server and SQL Server, supports teachers, students, academic services and the public. Católica - Porto's TELE is the product of this context. However, the consistency in the management of these environments at the institutional level is not easy to achieve, due to lack of critical and meaningful information. In this paper, we try to make a contribution on this topic. The analysis is focused on the institutional TELE, which in this case is supported by LCMS Blackboard.

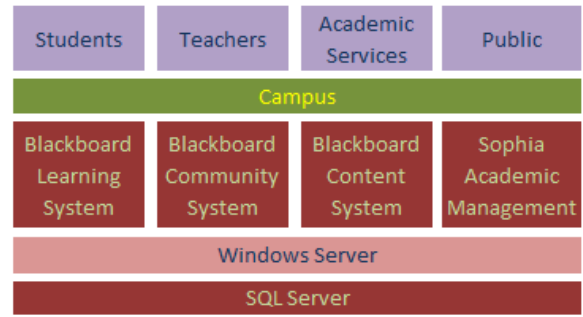

Figure 2. LCMS as a central element of the campus

Católica - Porto's TELE is currently a dynamic system. Between 17th October and 14th November 2011 there were 5,864 registered users of the system and 666 active CU. During this period, the maximum open sessions per hour hit 470. Fig. 2-5 show some data that proves the dynamics of the system.

The number of daily visits to the campus reaches, in the highest peaks, close to 4000 (Fig. 3), and the average time per visit is 7'53"; the number of visitors on the days of greater access exceeds 2000 (Fig. 4); there are many days when the peak of pages viewed/ day is located close to 60,000 (Fig. 5) and on average 16 pages are seen in each visit; the mobile access also reaches Fig. close to 150 hits on the days of higher peaks (Fig. 6).

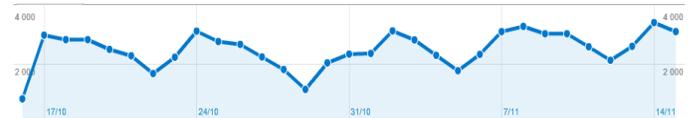

Figure 3. Daily visits to the campus

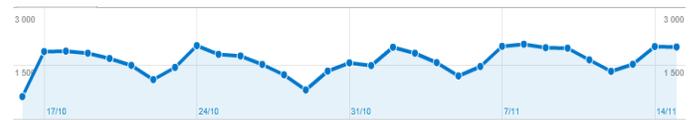

Figure 4. Visitors/day to the campus

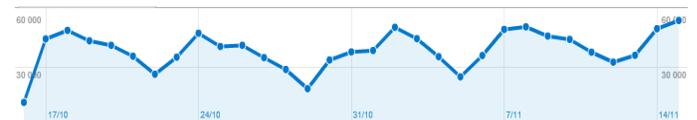

Figure 5. Pages seen/day

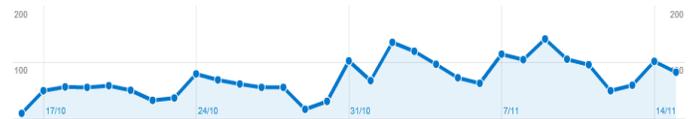

Figure 6. Mobile access

The financial investment made in acquiring the product and the services of the LCMS Blackboard and this dynamic of accesses justify the development of management tools that make available information on the integration of this TELE in the training process, guide educational policies and teacher training.

## III. METHODOLOGY

The methodological approach followed is part of the action research. Such research can be defined as follows: "Action research aims to contribute to both the practical concerns of people in an immediate problematic situation and to further the goals of social science simultaneously. Thus, there is a dual commitment in action research to study the system and concurrently to collaborate with members of the system in changing it in what is together regarded as a desirable direction" [11]. In Fig. 7 the main methodological steps followed are systematized.

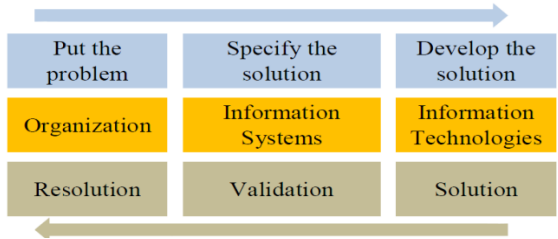

Figure 7. Methodological steps

1st *Put the problem*: The first step was to identify the problem at the organizational level. As already mentioned, there was a financial investment by the institution in the implementation of a TELE and the indicators provided by the Blackboard reports and via Google Analytics point to widespread use. It is important to justify the investment made and to understand the degree of integration of the LCMS in the formative process. The problem we face is to devise a management tool that delivers relevant information that can be aggregated according to the organizational plan of the institution.

2nd *Specify the solution*: Once the problem was identified [4], a solution, in which the way to use Information Systems (IS) is determinate and the architecture of the management tool is defined, was proposed in order to address the problem.

3rd *Develop the solution*: The implementation of the specified solution demands the improvement of the Blackboard reports. The inputs related to the definition of new information requirements, which respond to the needs of the institution and the users, were given by the investigators to the supplier of the service and to the representative of Blackboard. In a dialectical process, the provider of the services has made progressive approaches to the proposals presented by the researchers, which in turn have reoriented their solutions in an effort to reconcile their goals with the requirements of technological feasibility, presented by the engineers at Blackboard.

In addition to the reports of Blackboard, it was considered important to gather the views of users (teachers and students) on aspects related to the integration of the TELE in the teaching activity. This way, objective data on the exploitation of the features of TELE were compared with feedback from users. It has often been highlighted the importance of user involvement so that they understand and feel the usefulness of the IS [7]. In this way of conception of a management tool, this requirement is fully met in two ways: i) participation of researchers, who are also users, in the definition of the TELE automatic reports; ii) consultation of teachers and students who are TELE's users via a questionnaire.

TABLE I. PLANNING INFORMATION SYSTEMS

| Activities | Goals |
|---|---|
| Strategic Analysis | To identify the current situation of the organization and SI (Where are we?) |
| Strategic Definition | To identify the vision and strategies to achieve it (Where do we want to go?) |
| Strategic Implementation | To plan, oversee and review the strategy (How will we get there?) |

Each of these three methodological steps have correspondence in the three phases of the IS planning, synthesized by Varajão [7], as shown in Table I.

## IV. THE MANAGEMENT TOOL

Fig. 8 presents a simplified diagram of the main entities involved in the teaching process in Católica - Porto.

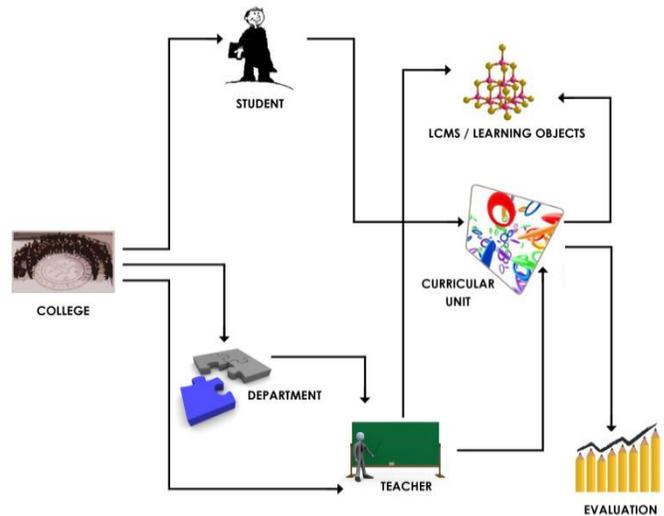

Figure 8. Simplified diagram of the main entities involved in the teaching process

Like many HEI, Católica - Porto has an organization on multiple levels: the university is divided into various colleges/ schools, that have under them several departments, in which the teachers are integrated (who can serve in different schools), who teach several CU (each CU is automatically created in the LCMS, virtual environment where part of the teaching activity takes place and where Learning Objects are made available). The institution's students are enrolled in certain CU and are automatically enrolled in these CU on LCMS (there is communication between the administrative management system and the Blackboard).

In the conception of the management tool, we tried to reflect this working model. In this context, the CU is the atom of information, from which the aggregation to higher levels is done, as it is shown in Fig. 9. One of the limitations identified in automatic reports from Blackboard was precisely the impossibility of making the aggregation of information across multiple levels [4].

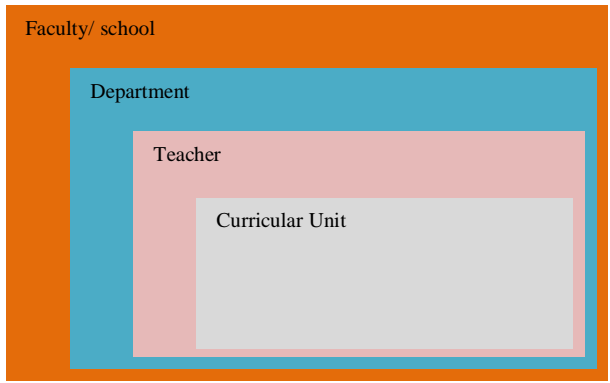

Figure 9. Aggregation of information reflecting the operating model

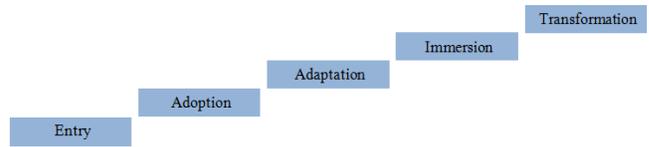

Figure 10. Levels of integration of the LCMS in the teaching and learning process

Another conclusion that was reached in the analysis phase was that the LCMS reports did not provide information on critical aspects to the understanding of the TELE's integration in teaching and learning. To address these limitations, seven dimensions that cover the main valences offered by the Blackboard were identified and indicators to characterize each of these dimensions were defined (table II). For each one of the indicators, metrics were established with five levels of integration in the formative process.

TABLE II. DIMENSIONS AND INDICATORS PROPOSED FOR THE AUTOMATIC LCMS REPORTS

| Dimensions | Indicators |
|---|---|
| Dynamics of accesses | Hits per week / active user |
| Information | Information relating to the CU through notices, messages, CU programme (or summary) and calendar |
| Synchronous Communication | Number of open forums and nr of posts/ active user |
| Asynchronous Communication | % of users who use one or more asynchronous communication tools within the LCMS |
| Digital Content | Number of digitally rich content (it is considered to be rich digital content, all that goes beyond text and static image. Example: podcasts, electronic presentations, games, animations,...) |
| Delivery of work | Use of features relating to the delivery of individual and group papers, progress monitoring of the work group, detection of plagiarism |
| Evaluation | Number of tests performed in LCMS |

Based on the indicators and metrics defined, a matrix with five levels of integration of the LCMS in the process of teaching and learning (Fig.10) was drawn up. This matrix of integration of technology in organizations and in the educational process through five stages of evolution is often present in studies [eg. 12, 13].

- *Entry*: Very low number of hits user/week. Lack of relevant information about the CU. No use or very little use of synchronous and asynchronous communication tools. Poor digital content. No delivery of works. No evaluation tests. The LCMS has a very limited impact on the teaching and learning process. It is possible to be successful at the CU without accessing the LCMS.

- *Adoption*: Low number of accesses user/week. Not much basic information about the CU. The synchronous and asynchronous communication tools are little explored. The presence of rich digital content is small, covering a small part of the key issues. The delivery of works via the platform is in its infancy, which limits the work of monitoring and detection of plagiarism. The assessment is only sporadically done and/or it is not important for the regulation of the study. The LCMS has limited impact, but it is visible in the process of teaching and learning. The student struggles to be successful in the CU without accessing the LCMS.

- *Adaptation*: The access to the CU is done regularly throughout the week. There is some basic information about the CU. The use of the synchronous and asynchronous communication tools is quite important in the construction of knowledge. The presence of rich digital content is visible, but they do not cover an important part of key issues. The delivery of works via the platform is sometimes associated with forms of communication, monitoring and detection of plagiarism. There are some key issues with assessment tests that are important for the regulation of the study. The LCMS has a clear impact on the teaching and learning process. It is extremely difficult for the student to be successful in CU without accessing the LCMS.

- *Immersion*: Access to CU is done on daily or almost daily basis. The majority of relevant information about the CU is available. The use of the synchronous and asynchronous communication tools is important in the construction of knowledge. There is digitally rich content that brings added value in comparison to printed material, covering most of the key themes of the CU. The delivery of works via the platform is usually associated with forms of communication, monitoring and detection of plagiarism. There are tests for most of the key issues that are important for the regulation of the study. The LCMS has a great impact on the teaching and learning process. The student cannot succeed without access to the LCMS.

- *Transformation*: The access to the CU is done daily or several times a day. All relevant information about the CU is available. The use of the tools synchronous and asynchronous communication is very important in the construction of knowledge. There is digitally rich content that brings great added value in comparison to printed material, covering most of the key themes of the CU. The delivery of works via the platform is always papers. There are tests for most of the key issues and they are very important for the regulation of the study. The LCMS is vital and has a transforming power in the process of teaching and learning.

In order to achieve this automatic placement via the LCMS reports, a back office system that allows the parameterization of the factors analyzed for each level of evolution was drawn up. To complement the information from the reports, which are translated into objective data on how the LCMS are used, it was considered important to ascertain the views of teachers and students about the same dimensions. Thus, the report data is compared with information from two questionnaires, which aims to understand the perspectives of teachers and students about key aspects of each dimension. Fig. 11 shows a possible form of representing the information: the radar charts.

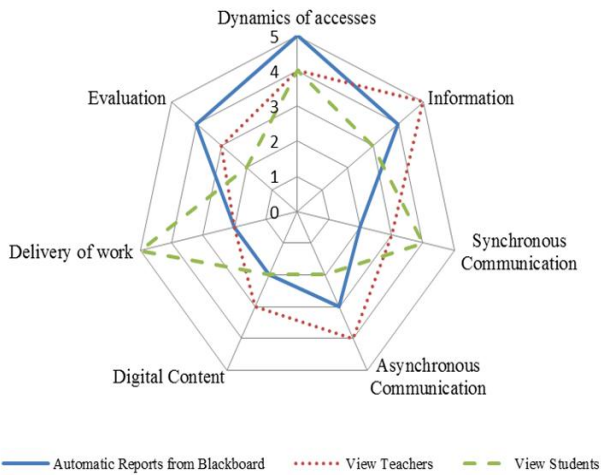

Figure 11. Radar Charts: Versatile forms of knowledge presentation

This type of chart, because of its versatility in knowledge representation, is often used in the analysis of organizational development and measurement of quality [14]. In an academic environment still in transition, where there is an increasing penetration of technology associated with pedagogical changes, radar charts are suitable and adaptable. In Fig. 11 the versatility of the radar chart is shown: appear represented in the same graph the dimensions of the TELE and the levels of integration of technology, where a comparative reading of data provenance can be made (Automatic reports from Blackboard, teacher's view, students' view).

The processing and the intelligent and versatile presentation of data are key features of management tools. Computer applications like OLAP (Online Analytical Processing) are useful as they offer these essential valences for decision-making [15].

The multidimensional OLAP functionality is based on structures called "cubes." The term "cube" is an analogy with the geometric object that implies three dimensions, but in real use, the OLAP cube can have more than three dimensions. The OLAP cube is comparable to a database, in which relationships between different dimensions and categories are established. It is a sophisticated technology that uses multidimensional structures to provide fast access to data for analysis. This organization facilitates the display of high-level of summaries. For the analysis of the TELE it can allow, for example, to extract data for each CU, for the dimensions in question, crossing automatic reports with the view of users.

The OLAP analysis system helps to organize data by many levels of detail (the information can be aggregated by CU, department, school/college or university), it also allows selecting and listing only the dimensions that we really want to consider in a given time. This strength allows conditional access to information, if this is the goal of the institution. In this case, each teacher will only have access to the information on the CU that he/she teaches, the coordinator of the department to all the CU of his/her department, the director of college/school to all the CU in the institution he/she manages, the Service Quality Management (SIGIQ) and the direction of Cattólica - Porto to all the information.

The data can also be selected by time periods or to view only a few variables. This way of presentation of information facilitates the interpretation of the results, since it reduces the entropy associated with high volumes of information.

In Fig. 12 the overview of the architecture proposed for evaluation of IS TELE is represented in a schematic form. The data comes from two sources: automatic reports from LCMS and the points of view of students and teachers (via questionnaire). The outputs of the management tool of information result in a matrix with five levels of integration of the LCMS in the process of teaching and learning. Currently, we are studying a way of providing information through an OLAP type application that allows a multidimensional analysis of data is under way.

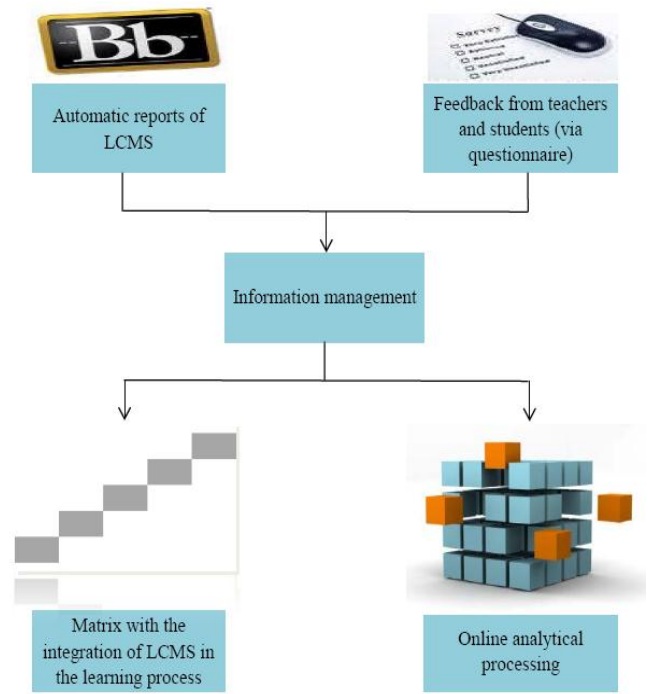

Figure 12. Architecture Overview of the management tool

## V. Conclusions

In this paper the process of conception of a tool for managing a TELE is described, and the three steps of the IS Planning model synthesized by Varajão [7] were followed:

*Strategic Analysis*: The delimitation of the problem was made. The shortcomings in the current IS when compared with the demands of information by the organizations in order to manage the TELE were identified.

*Strategic Definition*: In the conception of the management tool a response to the identified problems was given, by defining the dimensions of analysis (i - dynamic access; ii - CU information; iii - synchronous communication; iv - asynchronous communication; v – digital contents; vi - delivery of papers, vii - evaluation) and the way of aggregating data, so that an analysis at various levels, in accordance with the organizational plan of the institution, is possible.

*Solution Development*: Through an action research methodology, researchers (LCMS users) presented the inputs for the improvement of the automatic reports and, in a working process in partnership with the supplier of the technical services, successive approximations have been made to the solutions presented. In this dialectical process, the proposals have been adequate to the requirements of technological feasibility, presented by the company's technical representative of Blackboard. In developing the tool it is expected that this information will be complemented with data collected, via questionnaire, reflecting the views of users (teachers and students) on the various dimensions of the TELE.

The main advantages of the proposed management tool and the methodology used are: i) the involvement of users in the conception of the tool, a fact that enhances its usefulness, ii) the comparison between data from the LCMS automatic reports with the view of users, a factor that potentially increases the effectiveness of the tool as a management tool; iii) the possibility of aggregation for hierarchical levels, which reflect the organizational plan of the institution.

## VI. Future Work

As future work, it is predicted: i) to continue the improvement of the LCMS automatic reports and of the matrix to the position of the CU; ii) to develop a way to process and present OLAP data type, which articulates the results of the LCMS with the data from the questionnaires to teachers and students; iii) to integrate the OLAP component as a subsystem of quality management. Subsequently, it will be necessary to implement the management tool and carry out successive tests, in order to technically stabilize the system and refine the information output.

## References

[1] S. A. Ferreira and A. Andrade, "Models and instruments to assess Technology Enhanced Learning Environment in Higher Education," eLearning Papers, pp. 1-10, 2011.

[2] SNAHE, "E-learning quality: Aspects and criteria for evaluation of e-learning in higher education," Luntmakargatan2008.

[3] M. Rosenberg, Beyond e-learning. San Francisco: Pfeiffer, 2006.

[4] S. A. Ferreira and A. Andrade, "Aproximação a um modelo de análise da integração do LMS no processo formativo," in II CIDU 2011 - II Congreso Internacional da Docencia Universitária, Vigo, Spain, 2011.

[5] J. Kotter, Leading change. Massachusetts: Harvard Business School Press, 1996.

[6] J. Kotter and D. Cohen, The heart of change. Massachusetts: Harvard Business School Press, 2002.

[7] J. Varajão, A arquitetura da gestão de sistemas de informação, 2 ed. Lisboa: FCA - Editora Informática, 1998.

[8] R. Bogdan and S. Biklen, Investigação qualitativa em Educação. Porto: Porto editora, 1994.

[9] A. McCormack, The e-Skills Manifesto. A Call to Arms. Brussels, Belgium: European Schoolnet, 2010.

[10] S. A. Ferreira, C. Castro, and A. Andrade, "Cognitive Communication 2.0 in the Classroom - Resonance of an Experience in Higher Education," in 10th European Conference on e-Learning (ECEL 2011), University of Brigthon, Brigthon, United Kingdom, 2011, pp. 246-255.

[11] A. Dill, D. Gerwin, D. Mitchell, L. Russell, and B. Staley, Research Guidelines. Hadley - Massachusetts: WHSRP Planning Group, 2003.

[12] T. Teo and Y. Pian, "A model for Web adoption," Information & Management, vol. 41, pp. 457-468, 2004.

[13] T. Florida Center for Instructional. (2011). Technology Integration Matrix. Available: http://fcit.usf.edu/matrix/matrix.php

[14] D. Kaczynski, L. Wood, and L. Harding, "Using radar charts with qualitative evaluation: Techniques to assess change in blended learning," Active Learning in Higher Education, vol. 9, pp. 23-41, 2008.

[15] OLAP.COM. (2010). OLAP software and education wiki. Available: http://olap.com/w/index.php/OLAP_Education_Wiki

### Authors Profile

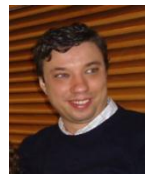

**Sérgio André Ferreira** Master in Sciences of Education, specialization in Education Computer Sciences. PhD student in Sciences of Education, specialization in Education Computer Sciences. His research is concerned with Technology Enhanced Learning Environments.

Webpage: http://www.sergioandreferreira.com/

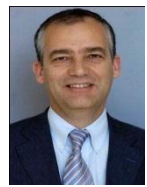

**António Andrade** Senior Lecturer at the School of Economics and Management of the Catholic Universidade Católica Portuguesa. PhD in Technologies and Information Systems, MSc in Information and Management, Director of the MSc in Information and Documentation. His research is concerned with Technology Enhanced Learning Environments. Webpage: http://www.porto.ucp.pt/feg/docentes/aandrade/